\documentclass{mn2e}
\usepackage{graphicx}

\title[Clumping in star-forming regions]
{Effects of clumping on temperature I:  externally heated clouds
}
\author[S. D. Doty et al.]
       {S. D. Doty$^1$, R. A. Metzler$^{1,2}$,
	M. L. Palotti$^{1,2}$ \\
        $^1$Department of Physics and Astronomy, Denison University,
	Granville, OH  43023, USA\\
	$^2$Department of Physics, University of Wisconsin-Madison,
	Madison, WI  53706-1390, USA 
	} 
\date{Accepted ; 
      Received  
      in original form }

\pagerange{\pageref{firstpage}--\pageref{lastpage}}
\pubyear{2005}

\begin{document}

\maketitle

\label{firstpage}

\begin{abstract}
We present a study of radiative transfer in dusty, clumpy
star-forming regions.   A series of self-consistent, 
three-dimensional, continuum radiative transfer models are constructed
for a grid of models parameterized by central luminosity,
filling factor, clump radius, and face-averaged optical depth.
The temperature distribution within the clouds is studied
as a function of this parameterization.  
Among our results, we find that: 
(a) the effective optical depth in clumpy regions is less than
in equivalent homogeneous regions of the same average
optical depth, leading to a deeper penetration of heating
radiation in clumpy clouds, and temperatures higher 
by over 60 per cent;  
(b)penetration of radiation is driven by the fraction of
open sky (FOS) -- which is a measure of the fraction of
solid angle along which no clumps exist; 
(c) FOS increases as
clump radius increases and as filling factor decreases; 
(d) for values of FOS $> 0.6-0.8$ the sky is sufficiently 
``open'' that the temperature distribution is relatively 
insensitive to FOS; 
(e) the physical process
by which radiation penetrates is preferentially through streaming
of radiation between clumps as opposed to diffusion through 
clumps; 
(f) filling factor always dominates the determination of the
temperature distribution for large optical depths, and for 
small clump radii at smaller optical depths; 
(g) at lower face-averaged
optical depths, the temperature distribution is most sensitive to 
filling factors of 1 - 10 per cent, in accordance with many observations;
(h) direct shadowing by clumps can be important for distances approximately
one clump radius behind a clump. 
 
\end{abstract}

\begin{keywords}
stars: formation -- infrared: stars -- ISM: clouds.
\end{keywords}

\section{Introduction}

An understanding of the star formation process requires an
understanding of the underlying density distribution in 
star forming regions.  In a direct sense, knowledge of
the density distribution can help distinguish between 
different potential dynamical scenarios for energy injection, 
collapse, fragmentation, and outflow.  In a more indirect
sense, the density distribution significantly affects our
ability to infer source properties through its influence
on thermal balance, chemistry, local emission, and
the processing of radiation (radiative transfer) between 
the emitting region and the observer.  

The density distribution is important for the dust as well
as the gas.  In particular, the dust forms the dominant source
of opacity to visible and infrared (IR) radiation, and is the
dominant source of IR continuum radiation.  Perhaps more 
importantly, the dust dominates thermal balance by direct
interaction with the radiation field (van de Hulst 1949), 
and through collisions with the gas (Greenberg 1971; 
Goldreich \& Kwan 1974).   As the problems of thermal balance
and radiative transfer are non-local, non-linear feedback problems,
comparison of detailed models with observations remain the
best choice of reliably inferring the source properties.

Source geometry is a significant problem in modeling 
star-forming regions.  While it is normal to assume
some geometric symmetry, such restrictions
are generally not realistic.  In particular, a wealth of observations
(e.g. Migenes et al. 1989; Dickman et al. 1990; Falgarone et al. 1991; 
Cesaroni et al. 1991; Marscher et al. 1993; Zhou et al. 1994; 
Plume et al. 1997; 
Shepherd et al. 1997) show that 
and fragmentation (see also Goldsmith 1996; Tauber 1996 and 
references therein).  
This is supported by dynamical models
(e.g. Truelove et al. 1998; Marinho \& Lepine 2000; 
Klapp \& Sigalotti 1998; Klessen 1997) which naturally produce
clumpy and fragmented structures.

Previous work on radiative transfer in dusty, clumpy 
environments has been undertaken (e.g. Hegmann \& Kegel 2003; 
Witt \& Gordon 1996; Varosi \& Dwek 1999; Boiss\'{e} 1990).
However, much of the previous work has concentrated on 
scattering (e.g. with application to reflection nebulae), 
the detailed methods of solution, and/or some of the fundamental
results such as ability of radiation to penetrate to apparently
high optical depths. In this paper we use a 3-D monte carlo
radiative transfer model to extend the previous work to 
a wider and different range of physical conditions.  In particular,
we construct a large grid of models in
an effort to better delineate and understand the effects of
clumpy media on radiative transfer.  By controlling and varying
the parameterization of the source, we attempt to disentangle 
some of the underlying physical causes of these effects.

In section two we describe the model.  We discuss the general
effects of clumping in section three.  In section four, 
we introduce the ``fraction of open sky'' (FOS), and
discuss the effects of number density, FOS, optical depth, 
filling factor, and clump size on the dust temperature
distribution.  We discuss the effects of shadowing in section five.
Finally, we draw conclusions in section six.

\section{Model}

\subsection{Monte carlo model and invariant parameters}
We have constructed detailed, self-consistent, three-dimensional
radiative transfer models through dust.  The model utilizes a
monte-carlo approach, combined with an approximate lambda
iteration to ensure true convergence even at high optical depths.  
The model has been tested against existing 1-D
(Egan, Leung, \& Spagna 1988) and 2-D (Spagna, Leung, \& Egan 1991) 
codes, and in modeling a 3-D source (Doty et al. 2005) with good success.  
Since the present study concentrates
upon opaque molecular clouds where far-infrared radiaton
should dominate both for external heating and emission, we 
ignore the effects of scattering.  Test cases in both one- and
multiple-dimensions show that scattering plays only a very small
role on the temperature distribution within the majority of the sources. 
 
Based upon the input parameters discussed below, we solve for the 
dust temperature and radiation field at each point in the
model cloud.  The computational volume is taken to be cubical 
of size 1 pc.
Each model utilized an $81 \times 81 \times 81$
cell grid, yielding a typical resolution of $\sim 3 \times 10^{16}$ cm.
The region is taken to be a two-phase medium consisting of high
density clumps, and a low density inter-clump medium.  The
clump/inter-clump density ratio is taken to be 
$n_{\mathrm{clump}}/n_{\mathrm{inter-clump}}=100$ from
observations (e.g. Bergin et al. 1996).   The external radiation field
is taken from interstellar radiation field (ISRF) compiled
by Mathis, Mezger, \& Panagia (1983).  Finally, we adopt the dust opacities
in column 5 of Table 1 
 of Ossenkopf \& Henning (1994), which have
been successful at fitting observations of both high-mass
(e.g., van der Tak et al. 1999, 2000) and low-mass
(e.g., Evans et al. 2001) regions of star formation.

\subsection{Model parameters}

We adopt a uniformly distributed interclump medium interspersed
with higher density clumps having
$n_{\mathrm{clump}}/n_{\mathrm{inter-clump}}=100$.
The clumps are randomly distributed within the computational 
volume.    The number of clumps, and the densities of the clumps
and interclump medium depend upon the filling factor ($f$), 
the clump radius ($r_{c}$), and the face-averaged optical depth 
($\bar{\tau}$).

The filling factor specifies the fraction of the volume at 
high density.  It is given by 
$f \equiv V_{\mathrm{clumps}}/V_{\mathrm{total}}$.
We adopt filling factors of $f=0.01$, $0.1$, and $0.3$
in accord with observations (e.g. Snell et al. 1984; 
Bergin et al. 1996; Carr 1987).  When all other parameters
are kept fixed, a higher $f$ corresponds to a larger
number of clumps, and a more nearly continuous dust
distribution.  Furthermore, due to optical depth
constraint (see $\bar{\tau}$ below), a larger $f$ also
corresponds to clumps and interclump medium of lower density.

We make the simplifying assumption that all non-overlapping
clumps are 
spheres of radius $r_{c}$. 
We choose
$r_{c} = 0.025$ pc, $0.05$ pc, and $0.1$ pc
in accord with observations (e.g. Carr 1987; Howe et al. 1993).
Again, the constraint on the optical depth implies that larger
clump radii yield smaller densities within the clumps.

The dust number densities are normalized by averaging the
optical depth over one entire ($81 \times 81$) 
face of the computational cube.  We call this the 
face-averaged optical depth, and denote it by 
$\bar{\tau}$.  This was done to simulate the
optical depth / column density that might be inferred by a very large beam, 
although it has the same effect as normalizing to the total
cloud mass.  The face-averaged optical depth is taken to be
$\bar{\tau} = 10$, and $100$, in keeping with observations 
from extinction studies (Lada, Alves, \& Lada 1999).

Finally, the models are specified by the strength of the 
internal radiation field, specified by the luminosity
of the central source, $L_{*}$.  We have
constructed a grid having $L_{*}= 0$L$_{\odot}$, $3$L$_{\odot}$, and
$300$L$_{\odot}$ to represent 
a starless core, a low-luminosity central object, and a high-luminosity
central object, respectively.  However, in this paper, we 
restrict our report to the starless cores only ($L_{*}=0$).  The
others will be presented in a forthcoming paper.

The ranges of parameters specified above led to a grid of 
54 models.  Each model is numbered by a four-digit integer
$I_{L}I_{f}I_{rc}I_{\bar{\tau}}$.  The integers, and their 
corresponding values are given in Table \ref{modelparameters}
for reference.
As an example, model 1221 has $L_{*}=0$, $f=0.01$, 
$r_{c}=0.025$ pc, and $\bar{\tau}=10$.

%_________________________BEGIN   ________________One column table
   \begin{table}
      \caption[]{Model input parameters}
         \label{modelparameters}
%     $$ 
%         \begin{array}{lllll}
         \begin{tabular}{rrrrr}
            \hline
            Number & $L_{*}$ & $f$ & $r_{c}$ & $\bar{\tau}$ \\
	     { }   & (L$_{\odot}$) & { } & (pc) & { } \\
            \hline
	     1 & $0$ & $0.01$ & $0.025$ & $10$  \\
	     2 & $3$ & $0.1$  & $0.05$  & $100$ \\
	     3 & $300$ & $0.3$ & $0.1$  & n/a  \\
	   
            \hline
     \end{tabular}\\
%         \end{array}
%     $$ 
   \end{table}
%____________________________END    ___________________ One column table

\subsection{Invariance with clump and photon randomization}

One concern with ``random'' clump distribution
and monte-carlo simulation is the reproducibility of results 
with different realizations 
of the clump distribution (i.e. different initial seeds in the 
clump generation), and different photon ray paths.  
As a test, we have considered 9 different initial seeds for the clump
distributions and photon paths respectively.  

To quantify the temperature
distribution, we calculate a spherical average temperature, 
given by 
\begin{equation}
\langle T(r) \rangle=\sum_{i}^{N} T_{i} n_{i} / \sum_{i}^{N} n_{i}.
\end{equation}
Here $N$ is the number of cells a distance between $r$ and
$r + \Delta r$ from the center, where $\Delta r$ is the size
of a single cell, and $T_{i}$ and
$n_{i}$ are the temperature and density of the cells respectively.  
In keeping with the viewpoint that the high density
clumps mostly affect the local radiation field, while the 
the low density medium best probes the radiation field, the average
temperature is taken only over the low-density cells.

The differences in $\langle T(r) \rangle$ between different
clump distribution realizations
are always much less than 1K, and are on average less than 
0.1K.  Based upon this (and direct comparison of 
later analyses) the specific realization of the density distribution
does not affect our conclusions.

%
%                                                One column figure
%------------------------------------------------- new & old rates 
   \begin{figure}
      \resizebox{\hsize}{!}{\includegraphics{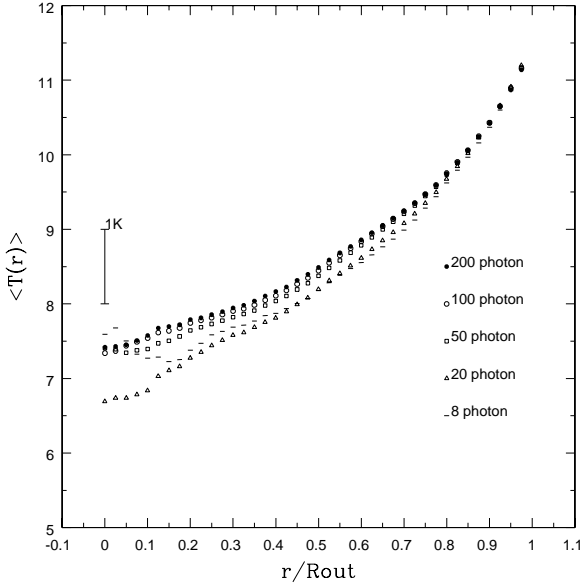}}
%%      \vspace{5cm}
      \caption[]{Spherical average temperature as a function of
      radial position for varying number of initial photons
      (ray pays) per cell face for the same random clumpy model.
              }
         \label{sphericalpn}
   \end{figure}
%%
%%______________________________________________________________

Likewise, we have modified the number and distribution of
incident photons to test for sufficient monte carlo coverage.
The effect of number of photons on for $\langle T(r) \rangle$  
is shown in Fig. \ref{sphericalpn}.
To be conservative, we adopt 100 photons per cell
face, at which point differences are less than 0.05K (1 per cent).

Similarly, varying the random distribution of
photon paths causes deviations of $< 0.1$K, confirming that the coverage  
of the cells by ray-paths is sufficient to 
concentrate on the consequences of clumping.

\section{Clumping:  general}

In this section we briefly discuss the general effects of
clumping on the dust temperature distribution.  For clarity,
we directly compare a clumpy, externally heated ($L_{*}=0$),
low-density ($f=0.1$), opaque ($\bar{\tau}=10$), model
having small clumps ($r_{c}=0.05$ pc) to one with a uniform
density distribution and the same optical depth and 
external heat source, so we can directly measure
the effects of clumping.  We have
done similar comparisons for all other models, with 
similar results.

%
%                                                One column figure
%------------------------------------------------- new & old rates 
   \begin{figure}
      \resizebox{\hsize}{!}{\includegraphics{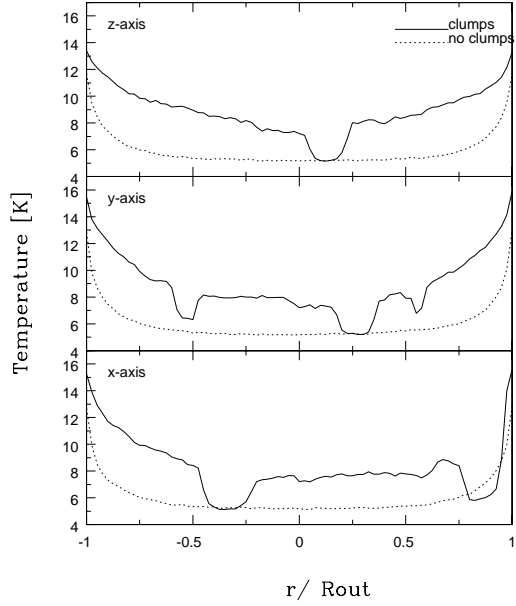}}
%%      \vspace{5cm}
      \caption[]{The temperature as a function of position within
      a clumpy (solid line) and non-clumpy (dotted line) low
      optical depth ($\bar{\tau}=10$) model, along the three
      principal axes.  
              }
         \label{tempclumpvnoclump}
   \end{figure}
%%
%%______________________________________________________________
%
%

%                                                One column figure
%------------------------------------------------- new & old rates 
   \begin{figure}
      \resizebox{\hsize}{!}{\includegraphics{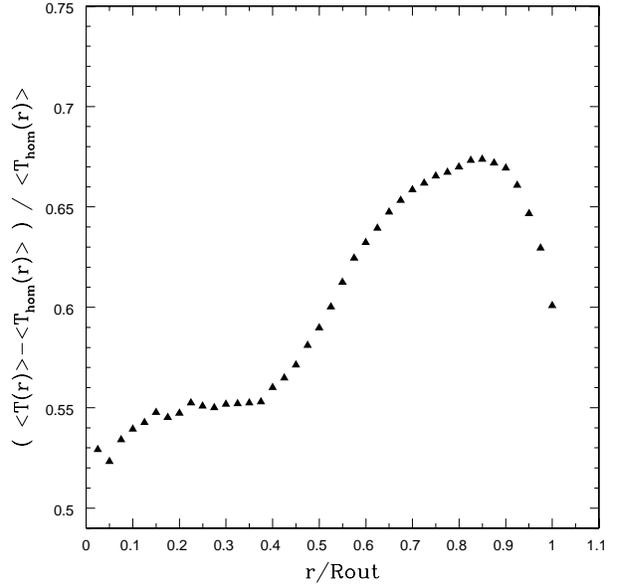}}
%%      \vspace{5cm}
      \caption[]{The fractional difference in $\langle T(r) \rangle$ 
      between a clumpy and non-clumpy externally heated ($L_{*}=0$),
      low optical depth ($\bar{\tau}=10$) model.
              }
         \label{clumpvnoclump}
   \end{figure}
%%
%%______________________________________________________________

The temperature distributions along the principal axes for 
a representative clumpy and equivalent homogeneous model 
are shown in Fig. \ref{tempclumpvnoclump}.  
To quantify the differences between the temperature 
distributions, in Fig. \ref{clumpvnoclump} we plot the fractional difference
in $\langle T(r) \rangle$ between these two representative models.
From these two figures, 
it is immediately obvious that the inclusion of clumps -- 
even for the same $\bar{\tau}$ -- changes the 
temperature structure significantly.   In particular, the
clumpy model experiences higher temperatures by up to 
50 per cent toward the center, and up to $\sim 65$ per cent
at intermediate radii.  
As a result, we conclude that clumping itself affects
the temperature distribution, even when the average source
mass or column density is held constant. 

This result confirms the previous finding of others
(e.g. Hegmann \& Kegel 2003; Witt \& Gordon 1996; 
Varosi \& Dwek 1999) that the effective optical depth in 
a clumpy medium is less than the homogeneous value.  
It also suggests that it is important to understand the
way in which the parameterization of the clumpy density
distribution can affect the dust temperature.  We address these
individually below.

\section{Clumping:  effects of parameters}

\subsection{Number density}
%
%                                                One column figure
%------------------------------------------------- new & old rates 
   \begin{figure}
      \resizebox{\hsize}{!}{\includegraphics{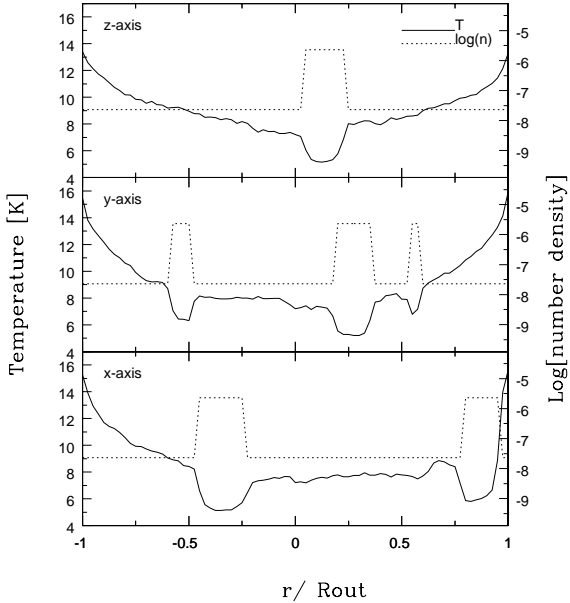}}
%%      \vspace{5cm}
      \caption[]{Distribution of temperature (solid lines, left-hand
      scale) and density (dotted lines, right-hand scale) for
      cuts along the x-, y-, and z-axes of model 1221.  
      Other models yield qualitatively 
      identical results.
              }
         \label{lintemp}
   \end{figure}
%%
%%______________________________________________________________

In Fig. \ref{lintemp} we plot the temperature (solid lines, left hand scale)
and dust number density (dotted lines, right hand scale) for
cuts along the x-, y-, and z-axes.  The clumps are signified by
the higher density regions.  As can be seen, the clumps are 
resolved, and appear to be of different sizes as the axes do not
penetrate all clumps along a diameter.  The ``wiggles'' and
$\sim 0.1 - 0.2$ K deviations in the 
temperature distribution are not simply indicative of the uncertainties
in the monte carlo calculation (see previous results, 
and Fig. \ref{tempclumpvnoclump}).  Instead, the
majority are dominated by local differences in radiation field
due to different amounts of blocking of radiation by the
surrounding clumps (see Sect. 5 for more discussion).

As can be seen in Fig. \ref{lintemp}, the dust
temperature is lower within the high density clumps than it is
within the lower-density interclump medium.  
%This is 
%expected, as the same incident flux in the case of a clump
%is spread-out over a larger number of dust grains.  As a result, 
%the energy density incident per grain is smaller for clumps than
%for the inter-clump medium, leading to lower temperatures. 
%    
Since the total emission by grains of absorption efficiency 
$\langle Q_{\mathrm{abs}} \pi a^2 \rangle \propto \nu^{\beta}$ goes as 
$T^{4 + \beta}$, one might expect a factor of 100 increase in
density to lead to a decrease in temperature by a factor 
$100^{1/(4 + \beta)}$.  Given that $\beta \sim 1.8$ in the FIR
for the adopted dust model, this is a decrease by a factor of $\sim 2.2$.  
For comparison, the average decrease within the clumps is
a factor of $1.5$ with a greatest decrease a factor of $1.6$.
This is due to the fact that the clump centers can be warmed by 
absorbing the radiation emitted from the clump edges, in effect
trapping radiation within the clumps (similar to the opaque 
centers of centrally heated envelopes, Doty \& Leung 1994).

Of even further interest is the fact that while the 
temperature decreases coincide with the locations of the clumps, 
the shape of the temperature profiles do not match the 
steepness of the clump/interclump interfaces.  This further
suggests that radiative transfer effects play a role.

\subsection{Effective optical depth}

%
%                                                One column figure
%------------------------------------------------- new & old rates 
   \begin{figure}
      \resizebox{\hsize}{!}{\includegraphics{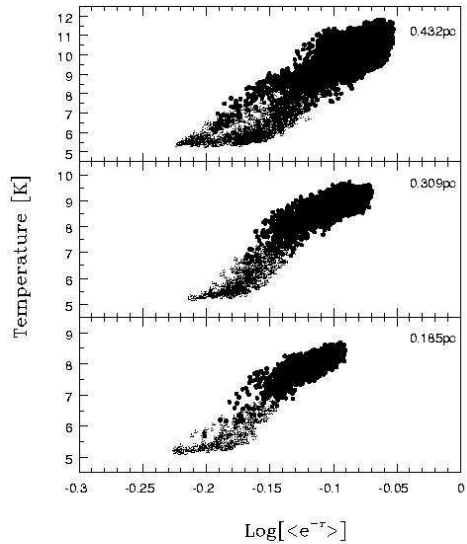}}
%%      \vspace{5cm}
      \caption[]{Temperature as a function of angle-averaged
      attenuation ($\langle e^{-{\tau}} \rangle $) for three distances from 
      source center ($R_{\mathrm{out}}/3$, $R_{\mathrm{out}}/2$,
      and $2R_{\mathrm{out}}/3$).  The solid symbols correspond to the
      temperature in the low density interclump medium, while the
      open symbols correspond to the high density clumps.  This
      result is for model 1221.  Other models
      are qualitatively identical.
              }
         \label{emtaubar3p}
   \end{figure}
%%
%%______________________________________________________________

Previous authors (e.g., Hegmann \& Kegel 2003; Witt \& Gordon 1996; 
Varosi \& Dwek 1999) identified the effective optical depth 
($- \ln[L_{\mathrm{AT}}/L_{*}]$), where $L_{\mathrm{AT}}$ is the
attenuated flux integrated over all solid angles and $L_{*}$ is the
unattenuated flux integrated over all solid angles or 
equivalently the average attenuation 
($\langle e^{- \tau} \rangle \equiv 
\int e^{- \tau} d{\Omega} / \int d{\Omega}$)
as a key measure of the ability of radiation to penetrate.
This is reasonable, as the fewer photons penetrating leads to cooler 
dust, which is
confirmed in Fig. \ref{emtaubar3p}, where we plot the dust
temperature at three different positions
along the x-axis as a function of $\langle e^{- \tau} \rangle$ for model 1221.

One question is left unanswered -- the 
physical process driving the angle-averaged attenuation
must be identified.
There are two possibilities:  
(1) the ability of radiation to penetrate is dominated by 
general attenuation by multiple clumps along a given line of sight and
attenuation by the low-density interclump medium; or 
(2) the radiation streams primarily through the ``holes'' between
clumps.  We consider these two possibilities in the following
subsection.

\subsection{Optical depth and fraction of open sky}

\subsubsection{Motivation and definition of $\langle \tau \rangle$ and FOS}
In order to answer the question of whether and how much geometry
matters in determining the local temperature/radiation field, we
consider two limiting cases.  These are a measure with limited
geometry information, the angle-averaged optical depth 
($\langle \tau \rangle$), and a measure which is dominated by 
geometry information, the fraction of open sky (FOS).  We consider
these seperately below.

If the radiation generally diffuses and is attenuated by 
multiple clumps or the low-density interclump medium, we might
expect to find a dependence of source properties on the 
optical depth averaged over all directions.  As a result, we
define the angle-averaged optical depth to be
\begin{equation}
\langle \tau \rangle \equiv \int \tau(\Omega) d \Omega / \int d \Omega.
\end{equation}

On the other hand, if the radiation penetrates mainly by
streaming through holes between clumps, then the source
properties should depend more significantly on the fraction of
the sky that is uncovered by the clumps.  Consequently, we also
define the fraction of open sky, FOS, to be
\begin{equation}
\mathrm{FOS} \equiv \int q d \Omega / \int d \Omega.
\end{equation}
Here $q = 1$ if there is no clump along the given line of sight
(i.e. if the sky is ``open'' in that direction), while
$q = 0$ if there is a clump along the given line of sight
(i.e. if the sky is ``closed'' in that direction).
Consequently, FOS=1 implies an open sky with no clumps, while
FOS=0 corresponds to a closed sky where all lines of sight
are blocked by clumps.

%
%                                                One column figure
%------------------------------------------------- new & old rates 
   \begin{figure}
%%      \resizebox{\hsize}{!}{\includegraphics{fostempcut1221.2.41.102.eps}}
      \resizebox{\hsize}{!}{\includegraphics{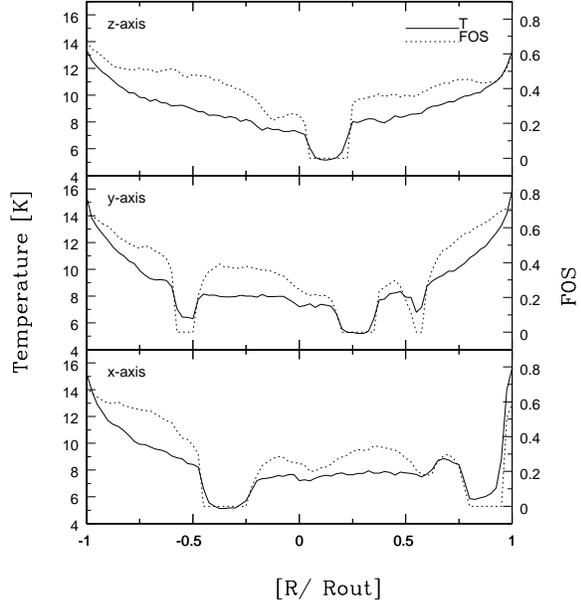}}
%%      \vspace{5cm}
      \caption[]{Distribution of temperature (solid lines, left-hand
      scale) and FOS (dotted lines, right-hand scale) for cuts along
      the x-, y-, and z-axes of model 1221.  
      Other models yield qualitatively identical results.
              }
         \label{fostempcut}
   \end{figure}
%%
%%______________________________________________________________

As a first test of the effect of FOS on temperature, in 
Fig. \ref{fostempcut} we plot the temperature (solid lines,
left-hand scale) and FOS (dotted lines, right-hand scale) for
cuts along the x-, y-, and z-axes, similar to
Fig. \ref{lintemp}.  There is a strong correlation
of dust temperature with FOS.  Interestingly, the
shape of the temperature distribution correlates with
FOS to a much higher degree than it anti-correlates with the dust number
density.  This strongly suggests that FOS plays
a key role, which we test in the following discussion.

\subsubsection{Effect of $ \langle \tau \rangle$ on temperature}

%
%                                                One column figure
%------------------------------------------------- new & old rates 
   \begin{figure}
      \resizebox{\hsize}{!}{\includegraphics{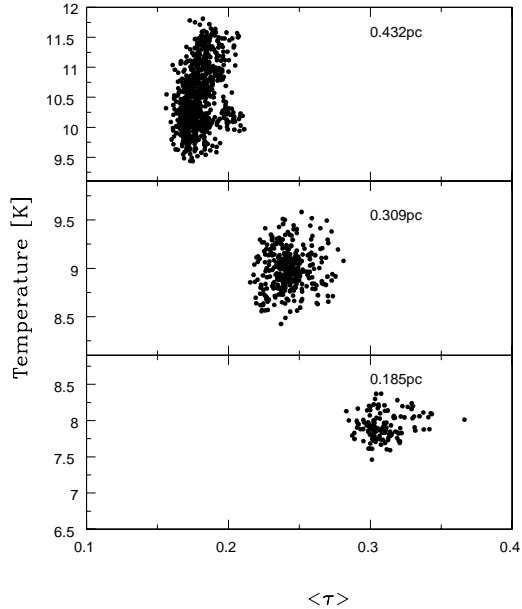}}
%%      \vspace{5cm}
      \caption[]{Temperature as a function of angle-averaged
      optical depth ($\langle \tau \rangle$) for three distances from 
      source center ($R_{\mathrm{out}}/3$, $R_{\mathrm{out}}/2$,
      and $2R_{\mathrm{out}}/3$), for a fixed FOS.
      This result is for model 1221.  Other models
      are qualitatively identical.
              }
         \label{taufosnew3p}
   \end{figure}
%%
%%___________________________________________________________

To test the role of angle-averaged optical depth, 
we consider the dependence of temperature on $\langle \tau \rangle$ for
a fixed FOS.  
In this way, we can isolate the effects of 
$\langle \tau \rangle$ from FOS.  
The FOS value here was chosen to maximize the
number of data points to provide for the best possible 
statistics.  
The resulting dependence of temperature
on $\langle \tau \rangle$ for three different radial positions is
shown in Fig. \ref{taufosnew3p}.  
We note that the chosen FOS range is small to make the result independent of
FOS, though moderate increases in the range lead to similar results.

%
%                                                One column figure
%------------------------------------------------- new & old rates 
   \begin{figure}
      \resizebox{\hsize}{!}{\includegraphics{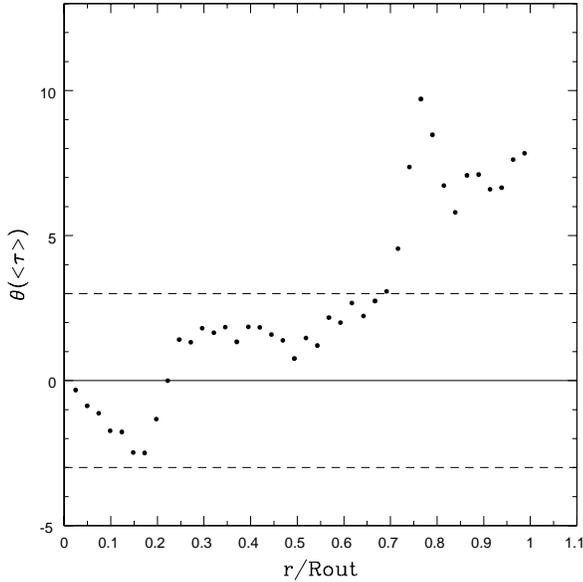}}
%%      \vspace{5cm}
      \caption[]{Slope of the best fit temperature - $\langle \tau \rangle$ 
      relationship
      divided by the uncertainty in the slope as a function of
      position in the source.  Notice the small correlation that
      exists, suggesting that $\langle \tau \rangle$ is of only minimal 
      importance in determining the temperature distribution.
      The dotted lines denote $\theta(\langle\tau\rangle)= \pm 3$ (i.e. 
      a 3$\sigma$ deviation from zero).
      This result is for model 1221.  Other models
      are qualitatively identical.
              }
         \label{taufosnewsig}
   \end{figure}
%%
%%__________________________________________________________________

There appears to be little correlation of temperature with
$\langle \tau \rangle$ at a given position.  
We quantify this by fitting the data distribution
at each radial distance with a best-fit straight line.  
To understand the significance of the slope / correlation, we
define a correlation parameter,
\begin{equation}
\theta({\langle \tau \rangle }) 
\equiv m({\langle \tau \rangle}) / \sigma_{m({\langle \tau \rangle})}.
\end{equation}
Here, $m(\langle{\tau}\rangle)$ is the slope of the best-fit line relating 
the temperature and $\langle \tau \rangle$, and 
$\sigma_{m({\langle \tau \rangle})}$ is the
uncertainty in that slope.
The results are shown
in Fig. \ref{taufosnewsig}.  We have also included
dashed lines at $m = \pm 3 \sigma_{m({\langle \tau \rangle})}$ as a guide.

To best understand the utility of this comparison, 
consider a case for which the temperature and $\langle \tau \rangle$ are
uncorrelated.  In this case, the fit between them 
should yield a zero slope, and thus $\theta({\langle \tau \rangle}) = 0$.
Likewise, a significant correlation should yield
$\theta > 3$ (i.e., a $3 \sigma$ result).
The results in Fig. \ref{taufosnewsig} are much more consistent with 
$\theta({\langle \tau \rangle})=0$ throughout much of the cloud, 
and only reach $\theta({\langle \tau \rangle})=3$ 
for $r>0.7R_{\mathrm{out}}$.  
Consequently, we conclude that there may exist only a weak 
correlation between the dust temperature and $\langle \tau \rangle$.

\subsubsection{Effect of FOS on temperature}

%
%                                                One column figure
%------------------------------------------------- new & old rates 
   \begin{figure}
      \resizebox{\hsize}{!}{\includegraphics{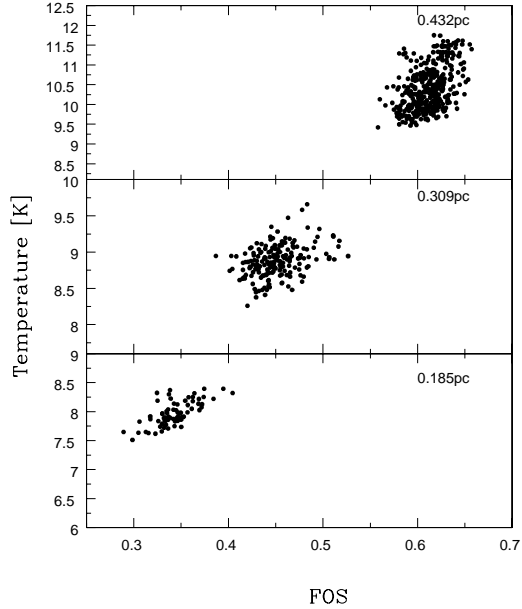}}
%%      \vspace{5cm}
      \caption[]{Temperature as a function of fraction
      of open sky (FOS)
      for three distances from 
      source center ($R_{\mathrm{out}}/3$, $R_{\mathrm{out}}/2$,
      and $2R_{\mathrm{out}}/3$), for a fixed $\langle \tau \rangle$.  
      This result is for model 1221.  Other models
      are qualitatively identical.
              }
         \label{fosnewtau3p}
   \end{figure}
%%
%%_______________________________________________________________

In a similar manner to $\langle \tau \rangle$, we consider the effect
of FOS on temperature.  In this case, we  
keep $\langle \tau \rangle$ fixed.
In analogy with before, $\langle \tau \rangle$ was
chosen to maximize the number of data points to ensure the
best possible statistics.
The resulting dependence of temperature
on FOS for three different radial positions is
shown in Fig. \ref{fosnewtau3p}.  

%
%                                                One column figure
%------------------------------------------------- new & old rates 
   \begin{figure}
      \resizebox{\hsize}{!}{\includegraphics{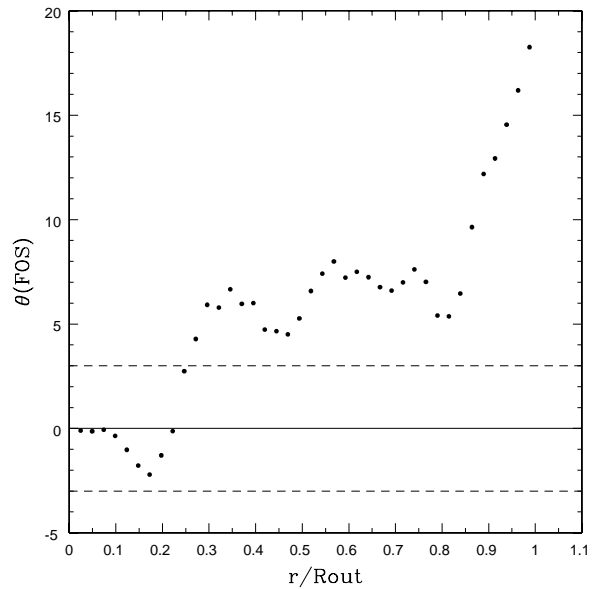}}
%%      \vspace{5cm}
      \caption[]{Slope of the best fit temperature - FOS relationship
      divided by the uncertainty in the slope as a function of
      position in the source.  Notice the large correlation that
      exists, suggesting that FOS has significance in
      determining the temperature distribution.
      This result is for model 1221.  Other models
      are qualitatively identical.
              }
         \label{fosnewtausig}
   \end{figure}
%%
%%__________________________________________________________________

Inspection of Fig. \ref{fosnewtau3p} suggests that there exists some
correlation of temperature with FOS.
In a similar manner to $\langle \tau \rangle$, 
we quantify this by fitting the data distribution
at each radial distance with a best-fit straight line.  
We again define a correlation parameter,
\begin{equation}
\theta({\mathrm{FOS}}) \equiv m({\mathrm{FOS}}) / \sigma_{m({\mathrm{FOS}})}.
\end{equation}
Here, $m({\mathrm{FOS}})$ is the slope of the best-fit line relating 
the temperature and FOS, and $\sigma_{m({\mathrm{FOS}})}$ is the
uncertainty in that slope.
The results are show 
in Fig. \ref{fosnewtausig}.  We have also included
dashed lines at $\pm 3 \sigma_{m({\mathrm{FOS}})}$ as a guide.

The results in Fig. \ref{fosnewtausig} are generally not
consistent with 
$\theta({\mathrm{FOS}})=0$, and lie at or above 
$\theta({\mathrm{FOS}})=3$ for a good deal of the cloud
($r > 0.25R_{\mathrm{out}}$).  
Consequently, we can say that there exists a significant 
correlation between the dust temperature and FOS.  

Finally, it is encouraging to note that the correlation between
temperature and FOS is positive.  This is expected, since
a higher FOS leads to more direct heating by the external
radiation field, and is evidenced by $m({\mathrm{FOS}}) > 0$.

\subsubsection{Extension and interpretation of $\langle \tau \rangle$, FOS results}

%_________________________BEGIN   ________________One column table
   \begin{table}
      \caption[]{Temperature dependence on FOS and $\langle \tau \rangle$.
      See text for discussion.}
         \label{fostautable}
         \begin{tabular}{rrrrr}
            \hline
            Model & $x_{\mathrm{max}}$ & 
   $x_{\mathrm{cut}}(\langle \tau \rangle)$ & 
   $x_{\mathrm{cut}}(\mathrm{FOS})$ & 
   $\overline{\theta(\mathrm{FOS})} / \overline{\theta(\langle \tau \rangle)}$\\
            \hline
	     1111 & 0.00 & 0.64 & 0.77 & n/a \\
	     1121 & 0.00 & 0.47 & 0.84 & n/a \\
	     1131 & 0.00 & 0.84 & 0.64 & n/a \\
	     1211 & 0.94 & 0.64 & 0.30 & 2.1 \\
	     1221 & 0.55 & 0.69 & 0.27 & $\infty$  \\
	     1231 & 0.00 & 0.74 & 0.99 & n/a \\
	     1311 & 0.99 & 0.10 & 0.30 & 1.4 \\
	     1321 & 0.97 & 0.97 & 0.22 & 4.1 \\
	     1331 & 0.86 & 0.94 & 0.92 & n/a \\
	     \\ 
	     1112 & 0.00 & 0.64 & 0.94 & n/a \\
	     1122 & 0.00 & 0.49 & 0.94 & n/a \\
	     1132 & 0.00 & 0.82 & 0.84 & n/a \\
	     1212 & 0.94 & 0.74 & 0.64 & 1.0 \\
	     1222 & 0.55 & n/a  & 0.99 & n/a \\
	     1232 & 0.00 & 0.79 & n/a  & n/a \\
	     1312 & 0.99 & 0.07 & 0.59 & 0.8 \\
	     1322 & 0.97 & 0.97 & 0.30 & 1.8 \\
	     1332 & 0.86 & 0.94 & n/a  & n/a \\
	       
            \hline
     \end{tabular}\\
%         \end{array}
%     $$ 
   \end{table}
%____________________________END    ___________________ One column table

%
%                                                One column figure
%------------------------------------------------- new & old rates 
   \begin{figure}
      \resizebox{\hsize}{!}{\includegraphics{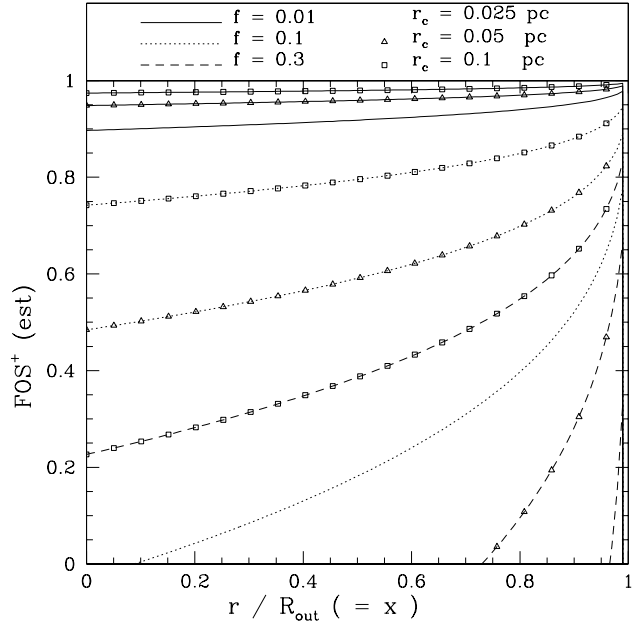}}
%%      \vspace{5cm}
      \caption[]{An estimate of the FOS in the outward facing
      hemisphere as a function of radial 
      position.
      The solid lines correspond to $f=0.01$, the dotted lines
      to $f=0.1$, and the dashed lines to $f=0.3$.  
      Lines without symbols correspond to $r_{c} = 0.025$ pc, 
      while those with triangles correspond to 
      $r_{c} = 0.05$ pc, and those with squares to 
      $r_{c} = 0.1$ pc.
      See text for details.
              }
         \label{plotfosplin}
   \end{figure}
%------------------------------------------------- new & old rates 

We can extend the discussion of FOS and $\langle \tau \rangle$
to the remainder of our 
family of models.  The results are summarized in Table \ref{fostautable}.
In this table, we specify the fractional radial position from 
cloud center as $x \equiv r/R_{\mathrm{out}}$. 

The results in Table \ref{fostautable} require an understanding of the
range over which FOS or $\langle \tau \rangle$ can have a significant
effect on the temperature distribution.  
In the table, $x_{\mathrm{max}}$ is the point beyond which,
FOS does not play a significant role on the temperature distribution.
We find (see Section 4.5) this occurs for $\mathrm{FOS} > 0.6-0.8$.   

To understand the origin and location of this region, consider a point 
a distance $r$ from the center of the region of radius $R_{\mathrm{out}}$.
The number of clumps exterior to $r$ is 
$N = f (R_{\mathrm{out}}/r)^{3}(1-x)$.  If these clumps are 
distributed about this point on the outward facing half of 
an equal volume sphere of radius 
$r_{\mathrm{eff}}=[4R_{\mathrm{out}}^{3}(1-x)/(4 \pi /3)]^{1/3}$, 
then the fraction of open sky in the outward direction is
$\mathrm{FOS}^{+} = 1 - Nr_{c}^2/2r_{\mathrm{eff}}^{2}$.
We plot the resulting estimate of FOS$^{+}$ as a function of position
in Fig. \ref{plotfosplin}.   As can
be seen, all models with a low filling factor (x1xx) have
FOS$^{+} > 0.8$, as does model (x23x).  For these models, FOS
is not expected to play a significant role in determining the
temperature distribution.  As a result, the radius up to which 
FOS should play a role ($\equiv x_{\mathrm{max}}$) becomes
zero for these models in Table \ref{fostautable}.
For other models, $0 < x_{\mathrm{max}} < 1$. 

To understand the comparative importance of FOS vs. 
$\langle \tau \rangle$, we consider the average deviation of
$\theta$ from zero for $x < x_{\mathrm{max}}$.  This
is reported in the last column of Table \ref{fostautable}.
For models in which $x_{\mathrm{max}} = 0$, such an average is
not possible, and the result is noted as ``n/a''.
For the case where only $\theta(\mathrm{FOS}) > 3$ for
$x < x_{\mathrm{max}}$, the only meaningful correlation of
temperature is with FOS, and so the result is noted as $\infty$.
In all other cases, the last column is set to 
$\overline{\theta(\mathrm{FOS})}/\overline{\theta(\langle \tau \rangle)}$, 
where the bar signifies an average over all positions for
which $\theta > 3$ and $x < x_{\mathrm{max}}$.
As can be seen by the last column of Table \ref{fostautable} the 
effect of FOS exceeds the effects of $\langle \tau \rangle$ in
determining the temperature.

Finally, one may consider the size of the region over which 
FOS and $\langle \tau \rangle$ are important.
To do this, in Table \ref{fostautable} we specify $x_{\mathrm{cut}}$,
defined to be the position beyond which $\theta > 3$.
%In the case presented (model 1221), we find 
%$x_{\mathrm{cut}}(\langle \tau \rangle) > x_{\mathrm{max}}$.
%On the other hand, $x_{\mathrm{cut}}(\mathrm{FOS}) = 0.27$.
Generally $x_{\mathrm{cut}}(\langle \tau \rangle) > 
x_{\mathrm{cut}}(\mathrm{FOS})$.   
At low optical depths, this is true whenever 
$x_{\mathrm{max}} \ne 0$.  For this set, the mean values for 
$x_{\mathrm{cut}}(\langle \tau \rangle) \sim 0.7$, and
$x_{\mathrm{cut}}(\mathrm{FOS}) \sim 0.4$.
This suggests that not only is FOS (and thus photon streaming) 
more important than $\langle \tau \rangle$, but that it is
also important deeper into the cloud than $\langle \tau \rangle$.
This is not a suprise, as openings
are a significantly more efficient method of carrying 
energy deep into the cloud.
At larger optical depths, the effect is not as strong due to the
fact that the less-dense interclump medium is capable of meaningful
attenuation by itself.   

The analysis above demonstrates that the angle-averaged
optical depth does not provide sufficient information for
describing a star-forming region.  This is seen in the fact
that the correlation of temperature with 
FOS is generally stronger than the correlation 
with $\langle \tau \rangle$, as well as the fact that the region
over which FOS is important is generally larger than the region
over which $\langle \tau \rangle$ is important.  
As a result, this suggests that 
indeed the ability of radiation to penetrate further into 
a clumpy source than a homogeneous source is due to the
streaming of radiation through holes.

%
%                                                One column figure
%------------------------------------------------- new & old rates 
   \begin{figure}
      \resizebox{\hsize}{!}{\includegraphics{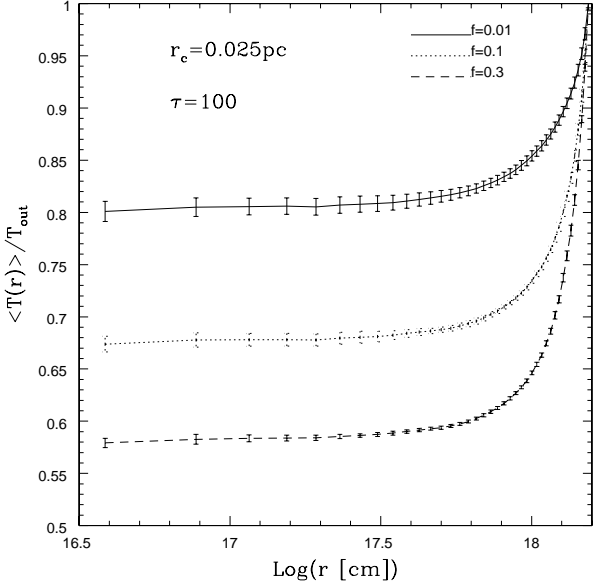}}
%%      \vspace{5cm}
      \caption[]{Ratio of local spherical average temperature to 
      the value at $r=R_\mathrm{out}$, 
      as a function of radial position, for a model with a clump size of
      $r_{c}=0.025$ pc, and $\bar{\tau}=100$.  
      The different lines correspond to 
      different filling factors:  $f=0.01$ (solid), $f=0.1$ (dotted),
      and $f=0.3$ (dashed).  The results are averaged over 9 different
      realizations of the clump distribution, and
      the error bars are the statistical uncertainties
      from the averaging.
              }
         \label{lowrc1t100}
   \end{figure}
%------------------------------------------------- new & old rates 

\subsection{Filling factor and clump radius}

In this subsection we investigate the effects of filling
factor and clump radius on the dust temperature distribution.
Since $f$ and $r_{c}$ both geometrically parameterizatize the 
clump distribution, we account for the range of
realizations in the actual clump distribution by averaging 
$\langle T(r) \rangle$ over nine different
realizations.  Finally, we note that in this subsection we present
the temperature contrast 
($\equiv \langle T(r_{\mathrm{in}}) \rangle / T_{\mathrm{out}}$), 
which minimizes the
effect of absolute density scaling between models.

%_________________________BEGIN   ________________One column table
   \begin{table}
      \caption[]{Temperature contrast
         $\langle T(r_{\mathrm{in}}) \rangle /
	  \langle T_{\mathrm{out}} \rangle$ : $\bar{\tau}=10$}
         \label{tinovertout10}
%     $$ 
%         \begin{array}{lllll}
         \begin{tabular}{rrrr}
            \hline
            $r_{c}$ (pc) & $f=0.01$ & $f=0.1$ & $f=0.3$ \\
            \hline
     $0.025$ & $0.69 \pm 0.01$ & $0.64 \pm 0.01$ & $0.56 \pm 0.01$  \\
     $0.05$  & $0.70 \pm 0.02$ & $0.67 \pm 0.07$ & $0.61 \pm 0.07$  \\
     $0.10$  & $0.70 \pm 0.04$ & $0.71 \pm 0.07$ & $0.72 \pm 0.05$  \\
            \hline
     \end{tabular}\\
%         \end{array}
%     $$ 
   \end{table}
%____________________________END    ___________________ One column table

\subsubsection{Filling factor}

Figure \ref{lowrc1t100} shows a representative 
variation in
the spherical average temperature distribution 
with filling factor 
($f$) 
for $r_c = 0.025$pc and $\bar{\tau}=100$.
The temperature distributions for other models in the grid are
qualitatively similar.  The ratio of inner to outer temperature
for the full grid (including uncertainties due to different
realizations) are shown in Tables \ref{tinovertout10} and
\ref{tinovertout100}.    

We see from these data that the role of filling
factor varies with clump radius.  
First, the dust temperature decreases with increasing $f$ as
expected, since a larger $f$ leads to a smaller
FOS, and thus less heating available to the lower 
density medium and lower temperatures.

Second, the temperature
distribution for the smallest filling factor
remains essentially unchanged as the clump size changes.  This is
due to the high FOS throughout, so that
each point is well-coupled to the external radiation
field, independent of the size of the clumps.

%_________________________BEGIN   ________________One column table
   \begin{table}
      \caption[]{Temperature contrast
         $\langle T(r_{\mathrm{in}}) \rangle /
	  \langle T_{\mathrm{out}} \rangle$ : $\bar{\tau}=100$}
         \label{tinovertout100}
%     $$ 
%         \begin{array}{lllll}
         \begin{tabular}{rrrr}
            \hline
             $r_{c}$ (pc) & $f=0.01$ & $f=0.1$ & $f=0.3$ \\
            \hline
     $0.025$ & $0.80 \pm 0.01$ & $0.67 \pm 0.01$ & $0.58 \pm 0.01$  \\
     $0.05$  & $0.80 \pm 0.01$ & $0.69 \pm 0.03$ & $0.55 \pm 0.02$  \\
     $0.10$  & $0.81 \pm 0.02$ & $0.71 \pm 0.03$ & $0.57 \pm 0.05$  \\
            \hline
     \end{tabular}\\
%         \end{array}
%     $$ 
   \end{table}
%____________________________END    ___________________ One column table

\begin{figure}
      \resizebox{\hsize}{!}{\includegraphics{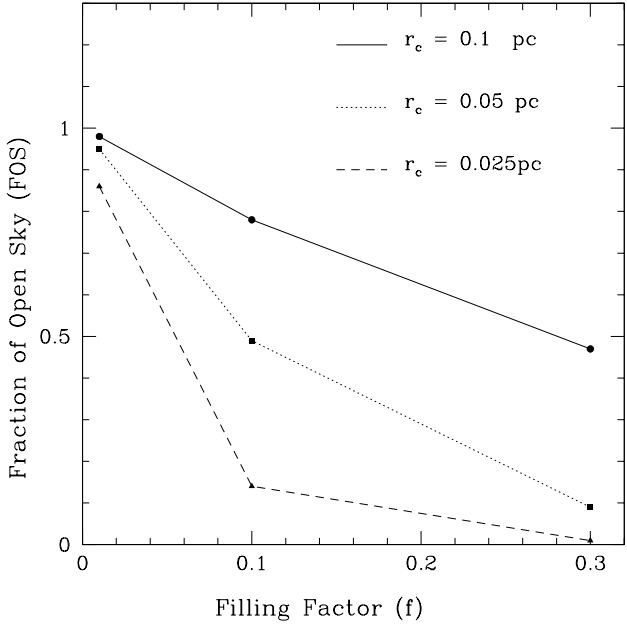}}
%%      \vspace{5cm}
      \caption[]{Fraction of open sky at the central point in
      the model cloud (FOS$^{c}$) as a function of filling
      factor for models $1xxx$ ($L_{*}=0$).  
      The lines correspond to different
      clump radii:  $r_{c}=0.025$ pc (solid), 
      $r_{c}=0.05$ pc (dotted), and $r_{c}=0.1$ pc (dashed).
      Note that the results are the same for all optical depths,
      for a given clump distribution.
      Results for different realizations of clump distributions
      are similar.
              }
         \label{plotfos}
   \end{figure}

Third, the effect of filling factor varies inversely with clump 
radius.  This is expected, as FOS increases with increasing $r_{c}$, 
and decreasing $f$.  To see this, consider the
simplifying case of viewing from the center of a sphere, 
in which spherical clumps of radius $r_c$ 
are randomly distributed a distance
$R_{\mathrm{out}}$ from the center, and in which no clumps overlap.  
This yields $\mathrm{FOS} = 1 - fR_{\mathrm{out}}/4r_{c}$.  
For $f=0.01 - 0.3$, this yields 
FOS $= 0.63-1$ for $r_{c} = 0.1$ pc, 
$0.25 - 1$ for $r_{c} = 0.05$ pc, and $\sim 0-1$ for 
$r_{c} = 0.025$ pc.  In reality  
FOS is influenced by two further competing factors -- clump  
overlap in projection (increases FOS), and clump distribution
in distance (decreases FOS).  The actual variation of 
FOS with model parameters is given in Fig. \ref{plotfos}.
The result roughly agrees with the estimate above.  The
lower trend of the actual results imply that 
clump overlap is not a significant effect.

%It is interesting to note in Fig. \ref{plotfos}
%that the range of FOS with $f$ is smallest for
%the largest clump radius ($r_{c}=0.1$ pc).  In this case, 
%the sky is mostly open for all filling factors, so that $f$ plays
%little if any role in the temperature determination.  On the
%other hand, for the smallest clump radii ($r_{c}=0.025$ pc), FOS
%varies considerably, and thus the temperature is strongly
%affected by filling factor.

These results are somewhat different for models with high optical depth,
$\bar{\tau} = 100$.  As confirmed in Table \ref{tinovertout100},
the temperature distribution at these high optical depths depends 
significantly on $f$, independent of $r_{c}$.

This independence of temperature on clump radius for large
optical depths can can be easily understood by considering the optical
depth of individual clumps for various models.  
For the ISRF of Mathis, Mezger, and Panagia (1983), roughly
1/2 of the total energy is contained in $\lambda < 25\mu$m, 
2/3 in $\lambda < 115\mu$m, and 3/4 in $\lambda < 360\mu$m. 
The optical depths of a clump at these wavelengths 
for $\bar{\tau}=10$ are, 
$1.1 < \tau_{\mathrm{clump}}(25\mu\mathrm{m}) < 67$, 
$0.1 < \tau_{\mathrm{clump}}(115\mu\mathrm{m}) < 6$, and
$0.01 < \tau_{\mathrm{clump}}(360\mu\mathrm{m}) < 0.8$.
On the other hand, 
for $\bar{\tau}=100$ the optical depths are ten times higher, namely, 
$11 < \tau_{\mathrm{clump}}(25\mu\mathrm{m}) < 670$, 
$1 < \tau_{\mathrm{clump}}(115\mu\mathrm{m}) < 60$, and
$0.1 < \tau_{\mathrm{clump}}(360\mu\mathrm{m}) < 8$.
This implies that the clumps for $\bar{\tau}=10$ are opaque to 
1/2 of the heating radiation, but transparent to the
remaining 1/2.  Consequently, $f$ (and FOS) can control
roughly 1/2 of the heating radiation.  However, for
$\bar{\tau}=100$, the clumps are opaque to 2/3 - 3/4 of the
heating radiation, meaning that $f$ (and FOS) can control
a greater amount of the energy that penetrates to 
a given depth.  As a result, it is not suprising to realize that
as the clumps become more opaque, filling factor and the ability 
of radiation to stream between the clumps becomes more important. 

%
%                                                One column figure
%------------------------------------------------- new & old rates 
%------------------------------------------------- new & old rates 
%

\subsubsection{Clump radius}

\begin{figure}
      \resizebox{\hsize}{!}{\includegraphics{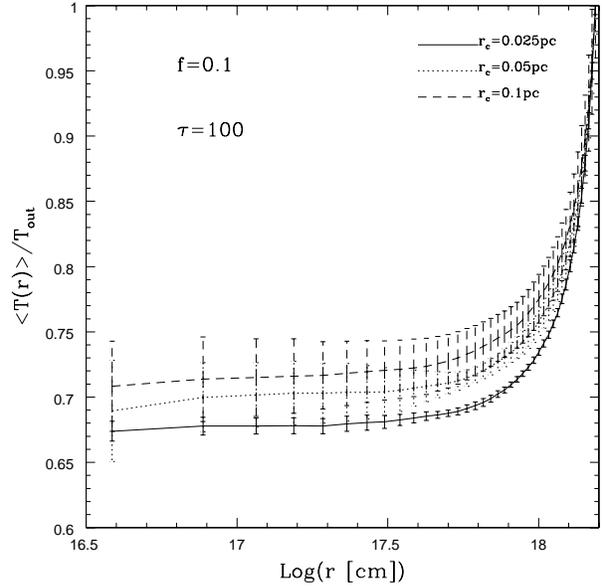}}
%%      \vspace{5cm}
      \caption[]{Ratio of local spherical average temperature to 
      the value at $r=R_\mathrm{out}$ as a function
      of radial position, for a model with a filling factor of
      $f=0.1$ pc, and $\bar{\tau}=100$.  The different lines correspond to 
      different clump radii:  $r_{c}=0.025$ (solid), 
      $r_{c}=0.05$ (dotted),
      and $r_{c}=0.1$ (dashed).  The results are averaged over 9 different
      realizations of the clump distribution, and
      the error bars are the statistical uncertainties
      from the averaging.
              }
         \label{lowf2t100}
   \end{figure}

Figure \ref{lowf2t100} shows a representative
variation in the spherical average temperature distribution with clump 
radius ($r_{c}$) for $f=0.1$ and $\bar{\tau}=100$.
Again, the temperature distributions in the
other models are qualitatively similar, and the temperature
contrast results are shown in Tables \ref{tinovertout10} \& 
\ref{tinovertout100}.

We see from these data that the role of clump radius 
varies with filling factor.  In comparison to the case of
filling factor, the smallest clumps yield the 
lowest temperatures, and the largest clumps the highest
temperatures, consistent with the FOS findings in Fig. \ref{plotfos}.

The independence of temperature with $r_c$ for $f=0.01$
is due to the large FOS.  In particular,  there exists a large number of open
lines of sight to any point, making clump variation unimportant.
Conversely, as $f$ increases, clump size becomes more important.
This is due to the fact that high filling factor models have a 
commensurately greater number of small clumps.  As discussed previously, 
many small clumps are more effective at covering the sky than 
fewer large clumps -- see Fig. \ref{plotfos}.  
Consequently, clump radius is an important factor in 
determining the temperature profile for high filling factors.  

%
%                                                One column figure
%------------------------------------------------- new & old rates 
%------------------------------------------------- new & old rates 

As before, the results differ for regions of higher optical
depth, $\bar{\tau}=100$.   
In Table \ref{tinovertout100} we can see the variation in 
the temperature contrast.   As discussed previously, the
increasing opacity of the clumps further amplifies the effect
of filling factor and FOS on the temperature distribution, 
leaving the effect of clump radius as unimportant.

\subsubsection{Relative regions of importance}

By combining the results of the two previous subsections
(e.g. Fig. \ref{plotfos} and Tables \ref{tinovertout10} \&
\ref{tinovertout100}), 
we can infer the regions of importance for filling 
factor and clump radius.  We first consider
the case of $\bar{\tau}=10$.
At these moderate optical depths, we see that 
clump radius only becomes important for
$f > 0.1$.  In the case of $f<0.1$, clump radius
essentially plays no role and is dominated by the fact that
the FOS is always high for such small filling factors.

On the other hand, we see that filling factor only
becomes important for $r_c < 0.05$ pc.  For 
$r_c > 0.05$ pc, the
temperature is dominated by the fact that the FOS is 
relatively insensitive to $f$, due to the large number of
holes left by scenarios with small numbers of large clumps. 

%
%                                                One column figure
%------------------------------------------------- new & old rates 
   \begin{figure}
      \resizebox{\hsize}{!}{\includegraphics{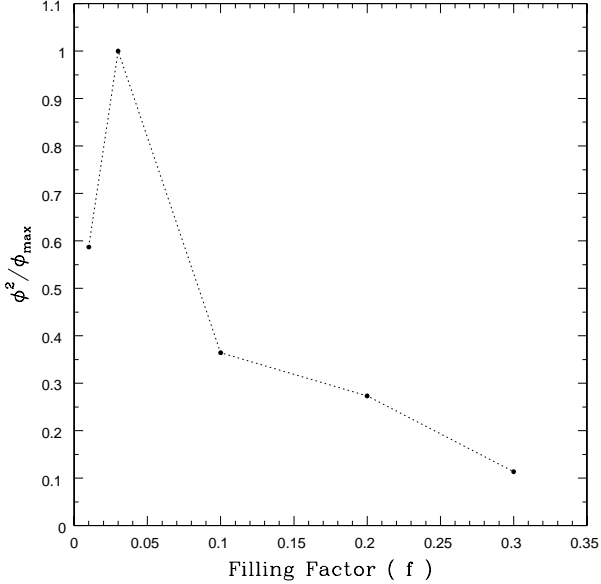}}
%%      \vspace{5cm}
      \caption[]{The deviation of the temperature distribution
      from an equivalent homogeneous model, as specified
      by $\phi(f)^2 / \phi_{\mathrm{max}}$ (see text) as 
      a function of the filling factor, for models
      with $r_{c}=0.5$ pc, and $\bar{\tau}=10$.
              }
         \label{phi2}
   \end{figure}
%------------------------------------------------- new & old rates 

It is interesting that the inferred values of $f$ from 
observations routinely fall in the range of a few per cent
(e.g. Hogerheijde, Jansen, van Dishoeck 1995; Snell et al. 1984; 
Mundy et al. 1986; Bergin 1996).  In light of the results
above, this may not be a suprise.  For smaller filling
factors, $f$ and $r_c$ do not play a significant role in the
temperature distribution.  On the other hand, for 
larger filling factors, the cloud is closer to homogeneous, 
and more sensitive to clump radius than $f$.  As a result,
the temperature distribution is most sensitive to changes
in $f$ for $f \sim 0.1$.  While many interpretations of
filling factor are not driven by continuum observations, 
these results are relevant since one of
the dominant thermal regulators for the gas is collisions
with the dust -- meaning that sensitivity to $T_{\mathrm{dust}}$ 
near $f \sim 0.1$ can lead to sensitivity in line
processes near $f \sim 0.1$.  

As a test, we have run additional models $f=0.03$, and
$f=0.2$ for the model numbers 1x21.  In order to 
quantify the deviation from the homogenous model, we have
calculated a parameter, 
$\phi^2 \equiv \frac{1}{N}\sum(\langle T(r_i) \rangle - 
\langle T_{\mathrm{hom}}(r_i) \rangle)^2 / \sigma(r)_{\mathrm{hom}}^2$.  
Here the sum is over radial positions, 
$\langle T_{hom}(r_i) \rangle$ is the spherical average
temperature for the homogenous model, and
$\sigma(r)_{\mathrm{hom}}$ is the uncertainty in the temperature
distribution in the homogenous model.  In Fig. \ref{phi2}
we plot the ratio of the $\phi(f)^2$ to the maximum value,
$\phi_{\mathrm{max}}^2$ as a function of filling factor.  Notice
that, indeed, the deviation from the homogeneous models
peaks for intermediate values of $f$.  In particular, the
greatest sensitivity to $f$ occurs in the range $f = 0.01 - 0.1$ -- 
a resultin accord with the ranges of filling factors commonly reported.

When the optical depth increases to $\bar{\tau}=100$, we find that
filling factor plays the dominant role over a much 
wider range.  As a result, the sensitivity to filling factor
near $f \sim 0.01-0.1$ for $\bar{\tau}=10$ is not an effect 
here.   At high optical depth, we also find that  
the clump radius is not important.  This is due to the
fact that $f$ has a larger effect on FOS than $r_c$ does, 
and that as optical depth increases, the clumps become
more opaque, thereby blocking a higher fraction of the
impinging radiation.

\subsection{Range of importance of FOS}

From the previous discussion, FOS is necessary
to describe the local radiation field and heating within a 
clumpy medium.  Here we attempt to infer the strength of the effect of FOS on
temperature.  While this has been addressed in
passing, here we collect and
further distill that information to help draw more general 
conclusions.

As seen in Section 4.3.4 and Fig. \ref{taufosnewsig}, the
greatest effect of $\theta(\mathrm{FOS})$ is at large radii.
We find that this is generally true across our grid of models.
As expected, these points tend to have the highest FOS.
Interestingly, however, the effect of FOS is on the situations where the
FOS is relatively small.  This can be inferred in three ways.

First, we note that the largest values of $\theta(\mathrm{FOS})$ occur
for the models that have the lowest FOS.  This can be indirectly
seen in that the models with the largest
$\overline{\theta(\mathrm{FOS})} / \overline{\theta(\langle \tau \rangle)}$
in Table \ref{fostautable} tend to have the larger filling factors.
As seen in Figs. \ref{plotfosplin} \& \ref{plotfos}, higher
$f$ corresponds to lower values of FOS.

We can roughly quantify this by saying that FOS is important
for those regions where FOS $\sim 0.5$ or so.
To begin to quantify this we can consider the temperature
distributions and deviations in Tables \ref{tinovertout10} \& 
\ref{tinovertout100}, together with the FOS
values in Fig. \ref{plotfos}.  In particular, we note that
meaningful differences in spherical average temperature 
occur between all models at the smallest clump radius,
that essentially no variation occurs
with filling factor for the largest clump radius,
and that meaningful variation may
occur as one changes filling factor at the intermediate
clump radius.  From Fig. \ref{plotfos}
we see that the largest clump radius has values of FOS $>0.5$.
On the other hand, for the smallest clump radius models, FOS varies
from $\sim 0.86$ to $\sim 0.25$ as one changes $f$.  Finally,
for the intermediate clump radius models, as one changes from
$f=0.01$ to $f=0.1$ FOS varies from $\sim 0.95$ to $\sim 0.5$,
with uncertain corresponding change in the temperature
distribution.  But, when the difference
between $f=0.01$ and $f=0.3$ is considered, there is a 
meaningful temperature change.  Taken together, this suggests that
variation in FOS for FOS $>0.5$ may not be as important as variations
in FOS when FOS $<0.5$ or so.

%
%                                                One column figure
%------------------------------------------------- new & old rates 
   \begin{figure}
      \resizebox{\hsize}{!}{\includegraphics{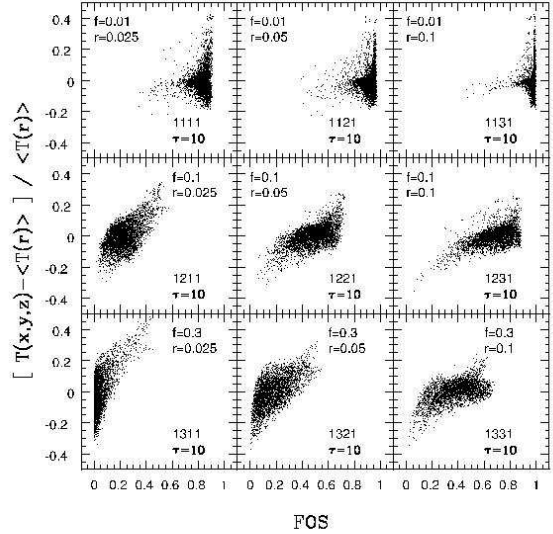}}
%%      \vspace{5cm}
      \caption[]{Deviation of temperature for a given point from the
      spherical average, as a function of the FOS for that point.  
      Each tenth point is plotted.  The grid of models having
      $\bar{\tau}$=10 (i.e. models 1xx1) are all plotted.  Notice the
      increased spread for FOS $> 0.6-0.8$, and the correlation
      for FOS $<0.6$ or so.
              }
         \label{plotranget10}
   \end{figure}
%%
%%______________________________________________________________

%
%                                                One column figure
%------------------------------------------------- new & old rates 
   \begin{figure}
      \resizebox{\hsize}{!}{\includegraphics{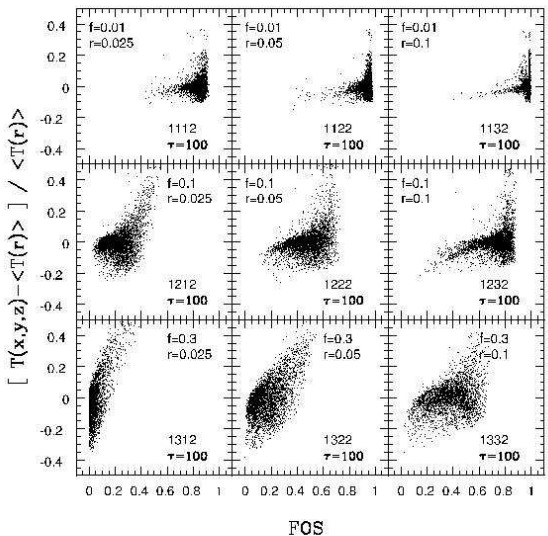}}
%%      \vspace{5cm}
      \caption[]{Deviation of temperature for a given point from the
      spherical average, as a function of the FOS for that point.  
      Each tenth point is plotted.  The grid of models having
      $\bar{\tau}$=100 (i.e. models 1xx2) are all plotted.  Notice the
      increased spread for FOS $> 0.6-0.8$, and the correlation
      for FOS $<0.6$ or so.
              }
         \label{plotranget100}
   \end{figure}
%%
%%______________________________________________________________
%

A more direct comparison suggests that FOS always plays some role,
but that a change in FOS becomes increasingly important in determining
the temperature for FOS $< 0.6$.  For FOS $>0.6-0.8$, the spread in 
temperatures increases.  To see this in Figs. \ref{plotranget10} \&
\ref{plotranget100} we have plotted the fractional deviation of 
the temperature of a point from the spherical average as a function of
the FOS at that point for all models in our grid.  The points
plotted cover all radii in the models.  It is clear by inspection that
FOS directly correlates with the temperature deviation.  Interestingly,
the correlation is strongest for FOS less than $\sim 0.6$.  On the
other hand, for FOS $>0.6-0.8$, there does not appear to be a direct
relationship between FOS and temperature deviation.  However, for
the larger values of FOS the spread in temperatures is indeed greater.

Taken together, this agrees well with the idea of streaming of
photons through holes in a clumpy medium.  For models with a small
average FOS, a position with a few extra lines of sight can
receive significant additional heating.  On the other hand, when the
FOS for a point is high, it is well-heated by the external radiation
field.  However, direct shadowing of a point by a nearby clump,
and attenuation by the (assumed) tenuous interclump medium can play roles.
In fact, for the high optical depth models ($\bar{\tau}=100$), the
ranges of temperture deviations are smaller than for low
optical depth models, due to the greater attenuation by the
interclump medium.

\section{Shadowing}

One further impact of clumps is to directly
shadow points behind them, leading to lower 
temperatures.  This effect can be seen directly 
in Fig. \ref{lintemp}.  In this externally heated
model, points just interior to clumps are decreased
in temperature relative to equivalent points not 
directly behind clumps.  Two important questions arise:
(1) over what length scale does this effect hold?, and
(2) is the cause a decrease in FOS for these points?

To answer the first question, we have considered the 
temperature distribution as a function of distance
behind clumps.  The temperature at these
points is consistently lower than the spherical average
by $5-25$ per cent.  To identify the shadowinglengthscale,
we determine the distance, $d$, behind the clump
at which the deviation is 1/2 the maximum.  
A fit of $d$ of the form
$d = \delta \times r_c$ yields $\delta =1.2 \pm 0.4$.  As a result, we infer 
shadowing is important for lengthscale of $\sim 1.2$ clump
radii behind a clump.   

To answer the second question -- the role of FOS -- we can
reconsider Fig. \ref{fostempcut}, which 
plots the FOS and temperature as functions of position
along three axes for a representative model.  Notice 
how well FOS correlates with temperature, especially 
in the shadowed regions behind clumps.  This is highly
suggestive that FOS is the determining factor in shadowing.
More quantitatively, we can consider the fraction of 
sky closed (FCS) by the clump.
Using the previous notation, the fraction of 
closed sky is $\mathrm{FCS} = r_{c}^{2}/(4d^2) = 1/4 \delta ^2$.  For 
$\delta=1.2$, this yields $\mathrm{FCS}_{\mathrm{clump}} = 0.17$, and
a corresponding FOS of $0.83$.  This is near
the limit of FOS$=0.8$ we inferred previously above which the
sky is sufficiently open for the temperature to be 
insensitive to FOS.  As a result, we conclude that FOS 
(and thus actual shadowing) is the important mechanism directly
behind a clump.

\section{Conclusions}
We have constructed a grid of three-dimensional continuum
radiative transfer models for clumpy star-forming regions, 
in order to better delineate and understand the effects of
clumping on radiative transfer.  Based upon this work, we
find that:

1.  The inclusion of clumps -- even for a constant total
mass / average optical depth -- can significantly change the 
temperature distribution within the cloud.  These differences
in temperature can be in excess of 60 per cent, and are
due to the lower effective optical depth for clumpy media
relative to equivalent homogeneous media (Sect. 3).

2.  The centers of clumps are warmer than would be expected
on energy density grounds due to radiation trapping (Sect. 4.1).

3.  The temperature distribution is driven by
the ability of radiation to penetrate, and is thus strongly
correlated with the angle-averaged attenuation 
$\langle e^{-\tau} \rangle$ (Sect. 4.2).

4.  While there exists an anti-correlation of temperature
with density (Sect. 4.1), the correlation with fraction
of open sky (FOS) is stronger (Sect. 4.2).

5.  We find only a weak correlation of dust temperature with
angle-averaged optical depth, $\langle \tau \rangle$ (Sect. 4.3.2).
On the other hand, there exists a significant correlation between
dust temperature and FOS (Sect. 4.3.3).  This is interpreted
as the dominance of streaming of radiation between clumps
over diffusion through them in determining the radiation field
(Sect. 4.3.4).

6.  The dependence of radiation penetration on FOS versus 
$\langle \tau \rangle$ is robust.  The stronger correlation with
FOS versus $\langle \tau \rangle$ extends throughout the grid of models
and for different realizations of clump distribution (Sect. 4.3.4).

7.  While $\langle \tau \rangle$ may be an effect 
near the cloud edges, FOS is more important deeper into the cloud 
(Sect. 4.3.4).

8.  At low face-averaged optical depths, $\bar{\tau}=10$, 
filling factor is more important for small clump radii than 
large clump radii.  At large
optical depths $\bar{\tau}=100$, filling factor is the
dominant effect for all situations (Sect. 4.4.1). 

9.  The effects of clump size are only important for the
largest filling factors ($f=0.3$) and lower optical depths
$\bar{\tau}=10$.  It is unimportant for lower filling factors
or larger optical depths, $\bar{\tau}=100$ (Sect. 4.4.2).

10.  For lower face-average optical depths, $\bar{\tau}=10$, 
filling factor is most important in determining the temperature
distribution for $f=0.01-0.1$, in accordance with most observations.
For very opaque clouds with $\bar{\tau}=100$, filling factor is
important over a larger range (Sect. 4.4.3).

11. FOS increases as clump radius increases and as filling factor
decreases (Sect. 4.4).  

12.  The variation of temperature with FOS is more significant
in cases of small FOS (high filling factor or small clump radii),
while $\langle \tau \rangle$ is relatively unimportant (Sect. 4.5).

13. For FOS $> 0.6 - 0.8$ the sky is sufficiently open that
there is little dependence of temperature on FOS (Sect. 4.5).

14.  Clumps can directly shadow the regions behind them.  On
average, this regime extends to distance $\sim 1.2$ times the clump
radius behind the clump, where the clump only subtends a small
fraction of the sky (Sect. 5).

\section*{Acknowledgements}
      We are grateful to Sheila Everett, Lee Mundy, and
      Dan Homan for thoughtful comments and interesting discussions,
      and the referee whose comments significantly improved the
      presentation.
      This work was partially supported under
      a grant from The Research Corporation (SDD), and Battelle.

\label{lastpage}

\end{document}